\documentclass[aps,prl,reprint,superscriptaddress]{revtex4-1}

\usepackage{graphicx}
\usepackage{amsmath}
\usepackage{longtable}
\usepackage{lineno}
\usepackage{gensymb}

\begin{document}


\title{First Measurement of the Total Neutron Cross Section on Argon Between 100 and 800 MeV}




\author{B.~Bhandari} 
\affiliation{Department of Physics, University of Houston, Houston, TX 77204, USA}
\author{J.~Bian} 
\affiliation{Department of Physics and Astronomy, University of California, Irvine, CA 92697, USA}
\author{K.~Bilton} 
\affiliation{Department of Physics, University of California, Davis, CA 95616 }
\author{C.~Callahan} 
\affiliation{Department of Physics and Astronomy, University of Pennsylvania, Philadelphia, PA 19104, USA}
\author{J.~Chaves}
\email[]{jchaves@sas.upenn.edu}
\affiliation{Department of Physics and Astronomy, University of Pennsylvania, Philadelphia, PA 19104, USA}
\author{H.~Chen}
\affiliation{Brookhaven National Laboratory, Upton, NY 11973, USA}
\author{D.~Cline}
\thanks{Deceased}
\affiliation{Department of Physics and Astronomy, University of California, Los Angeles, CA 90095, USA}
\author{R.~L.~Cooper}
\affiliation{Department of Physics, New Mexico State University, Las Cruces, NM 88003, USA}
\author{D.~Danielson} 
\affiliation{Department of Physics, University of California, Davis, CA 95616 }
\author{J.~Danielson} 
\affiliation{Los Alamos National Laboratory, Los Alamos, NM 87545, USA}
\author{N.~Dokania} 
\affiliation{Department of Physics and Astronomy, Stony Brook University, Stony Brook, NY 11794, USA}
\author{S.~Elliott} 
\affiliation{Los Alamos National Laboratory, Los Alamos, NM 87545, USA}
\author{S.~Fernandes}
\affiliation{Department of Physics and Astronomy, University of Alabama, Tuscaloosa, AL 35487, USA}
\author{S.~Gardiner}
\affiliation{Department of Physics, University of California, Davis, CA 95616 }
\author{G.~Garvey} 
\affiliation{Los Alamos National Laboratory, Los Alamos, NM 87545, USA}
\author{V.~Gehman} 
\affiliation{Lawrence Berkeley National Laboratory, Berkeley, CA 94720, USA}
\author{F.~Giuliani} 
\affiliation{Department of Physics and Astronomy, University of New Mexico, Albuquerque, NM 87131, USA}
\author{S.~Glavin} 
\affiliation{Department of Physics and Astronomy, University of Pennsylvania, Philadelphia, PA 19104, USA}
\author{M.~Gold} 
\affiliation{Department of Physics and Astronomy, University of New Mexico, Albuquerque, NM 87131, USA}
\author{C.~Grant}
\affiliation{Department of Physics, Boston University, Boston, MA 02215, USA}
\author{E.~Guardincerri} 
\affiliation{Los Alamos National Laboratory, Los Alamos, NM 87545, USA}
\author{T.~Haines} 
\affiliation{Los Alamos National Laboratory, Los Alamos, NM 87545, USA}
\author{A.~Higuera} 
\affiliation{Department of Physics, University of Houston, Houston, TX 77204, USA}
\author{J.~Y.~Ji} 
\affiliation{Department of Physics and Astronomy, Stony Brook University, Stony Brook, NY 11794, USA}\author{R.~Kadel}
\affiliation{Lawrence Berkeley National Laboratory, Berkeley, CA 94720, USA}
\author{N.~Kamp} 
\affiliation{Los Alamos National Laboratory, Los Alamos, NM 87545, USA}
\author{A.~Karlin} 
\affiliation{Department of Physics and Astronomy, University of Pennsylvania, Philadelphia, PA 19104, USA}
\author{W.~Ketchum} 
\affiliation{Los Alamos National Laboratory, Los Alamos, NM 87545, USA}
\author{L.~W.~Koerner} 
\affiliation{Department of Physics, University of Houston, Houston, TX 77204, USA}
\author{D.~Lee} 
\affiliation{Los Alamos National Laboratory, Los Alamos, NM 87545, USA}
\author{K.~Lee} 
\affiliation{Department of Physics and Astronomy, University of California, Los Angeles, CA 90095, USA}
\author{Q.~Liu} 
\affiliation{Los Alamos National Laboratory, Los Alamos, NM 87545, USA}
\author{S.~Locke} 
\affiliation{Department of Physics and Astronomy, University of California, Irvine, CA 92697, USA}
\author{W.~C.~Louis} 
\affiliation{Los Alamos National Laboratory, Los Alamos, NM 87545, USA}
\author{A.~Manalaysay} 
\affiliation{Department of Physics, University of California, Davis, CA 95616 }
\author{J.~Maricic} 
\affiliation{Department of Physics and Astronomy, University of Hawaii at Manoa, Honolulu, HI 96822, USA}
\author{E.~Martin} 
\affiliation{Department of Physics and Astronomy, University of California, Los Angeles, CA 90095, USA}
\author{M.~J.~Martinez}
\affiliation{Los Alamos National Laboratory, Los Alamos, NM 87545, USA}
\author{S.~Martynenko} 
\affiliation{Department of Physics and Astronomy, Stony Brook University, Stony Brook, NY 11794, USA}
\author{C.~Mauger}
\affiliation{Department of Physics and Astronomy, University of Pennsylvania, Philadelphia, PA 19104, USA}
\author{C.~McGrew} 
\affiliation{Department of Physics and Astronomy, Stony Brook University, Stony Brook, NY 11794, USA}
\author{J.~Medina} 
\affiliation{Los Alamos National Laboratory, Los Alamos, NM 87545, USA}
\author{P.~J.~Medina}
\thanks{Deceased}
\affiliation{Los Alamos National Laboratory, Los Alamos, NM 87545, USA}
\author{A.~Mills} 
\affiliation{Department of Physics and Astronomy, University of New Mexico, Albuquerque, NM 87131, USA}
\author{G.~Mills}
\thanks{Deceased}
\affiliation{Los Alamos National Laboratory, Los Alamos, NM 87545, USA}
\author{J.~Mirabal-Martinez} 
\affiliation{Los Alamos National Laboratory, Los Alamos, NM 87545, USA}
\author{A.~Olivier}
\affiliation{Department of Physics and Astronomy, Louisiana State University, Baton Rouge, LA 70803, USA}
\author{E.~Pantic} 
\affiliation{Department of Physics, University of California, Davis, CA 95616 }
\author{B.~Philipbar}
\affiliation{Department of Physics and Astronomy, University of New Mexico, Albuquerque, NM 87131, USA}
\author{C.~Pitcher} 
\affiliation{Department of Physics and Astronomy, University of California, Irvine, CA 92697, USA}
\author{V.~Radeka} 
\affiliation{Brookhaven National Laboratory, Upton, NY 11973, USA}
\author{J.~Ramsey} 
\affiliation{Los Alamos National Laboratory, Los Alamos, NM 87545, USA}
\author{K.~Rielage} 
\affiliation{Los Alamos National Laboratory, Los Alamos, NM 87545, USA}
\author{M.~Rosen} 
\affiliation{Department of Physics and Astronomy, University of Hawaii at Manoa, Honolulu, HI 96822, USA}
\author{A.~R.~Sanchez}
\affiliation{Los Alamos National Laboratory, Los Alamos, NM 87545, USA}
\author{J.~Shin} 
\affiliation{Department of Physics and Astronomy, University of California, Los Angeles, CA 90095, USA}
\author{G.~Sinnis} 
\affiliation{Los Alamos National Laboratory, Los Alamos, NM 87545, USA}
\author{M.~Smy}
\affiliation{Department of Physics and Astronomy, University of California, Irvine, CA 92697, USA}
\author{W.~Sondheim} 
\affiliation{Los Alamos National Laboratory, Los Alamos, NM 87545, USA}
\author{I.~Stancu}
\affiliation{Department of Physics and Astronomy, University of Alabama, Tuscaloosa, AL 35487, USA}
\author{C.~Sterbenz} 
\affiliation{Los Alamos National Laboratory, Los Alamos, NM 87545, USA}
\author{Y.~Sun}
\affiliation{Department of Physics and Astronomy, University of Hawaii at Manoa, Honolulu, HI 96822, USA}
\author{R.~Svoboda} 
\affiliation{Department of Physics, University of California, Davis, CA 95616 }
\author{C.~Taylor} 
\affiliation{Los Alamos National Laboratory, Los Alamos, NM 87545, USA}
\author{A.~Teymourian} 
\affiliation{Department of Physics and Astronomy, University of California, Los Angeles, CA 90095, USA}
\author{C.~Thorn}
\affiliation{Brookhaven National Laboratory, Upton, NY 11973, USA}
\author{C.~E.~Tull}
\affiliation{Lawrence Berkeley National Laboratory, Berkeley, CA 94720, USA}
\author{M.~Tzanov}
\affiliation{Department of Physics and Astronomy, Louisiana State University, Baton Rouge, LA 70803, USA}
\author{R.~G.~Van de Water} 
\affiliation{Los Alamos National Laboratory, Los Alamos, NM 87545, USA}
\author{D.~Walker}
\affiliation{Department of Physics and Astronomy, Louisiana State University, Baton Rouge, LA 70803, USA}
\author{N.~Walsh}
\affiliation{Department of Physics, University of California, Davis, CA 95616 }
\author{H.~Wang}
\affiliation{Department of Physics and Astronomy, University of California, Los Angeles, CA 90095, USA}
\author{Y.~Wang} 
\affiliation{Department of Physics and Astronomy, University of California, Los Angeles, CA 90095, USA}
\author{C.~Yanagisawa}
\affiliation{Department of Physics and Astronomy, Stony Brook University, Stony Brook, NY 11794, USA}
\author{A.~Yarritu}
\affiliation{Los Alamos National Laboratory, Los Alamos, NM 87545, USA}
\author{J.~Yoo}
\affiliation{Department of Physics, University of Houston, Houston, TX 77204, USA}


\collaboration{CAPTAIN Collaboration}
\noaffiliation

\date{\today}

\begin{abstract}
We report the first measurement of the neutron cross section on argon in the energy range of 100-800 MeV. The measurement was obtained with a 4.3-hour exposure of the Mini-CAPTAIN detector to the WNR/LANSCE beam at LANL. The total cross section is measured from the attenuation coefficient of the neutron flux as it traverses the liquid argon volume. A set of 2,631 candidate interactions is divided in bins of the neutron kinetic energy calculated from time-of-flight measurements. These interactions are reconstructed with custom-made algorithms specifically designed for the data in a time projection chamber the size of the Mini-CAPTAIN detector.  The energy averaged cross section is  $0.91 \pm{} 0.10~\mathrm{(stat.)} \pm{} 0.09~\mathrm{(sys.)}~\mathrm{barns}$.  A comparison of the measured cross section is made to the GEANT4 and FLUKA event generator packages, where the energy averaged cross sections in this range are $0.60$ and $0.68$ $\mathrm{barns}$ respectively.
\end{abstract}


\maketitle



Neutron interactions in liquid argon have not been studied for neutron kinetic energies above 50 MeV\cite{old_neutron}. This poses a problem for current and future liquid argon neutrino experiments\cite{DUNE}\cite{MicroBOONE} given the expected resolution to measure parameters such as the CP violating phase and determine the mass hierarchy in the neutrino sector. These measurements rely on knowing precisely the energy of the incoming neutrino. 
The energy of individual neutrino interactions is not known \textit{a priori} and must be reconstructed from the resulting particles of the interaction. 
In most cases, the neutrino interaction produces a charged lepton and additional hadrons. These hadrons will sometimes interact with other nucleons before they leave the nucleus. 
The final state hadronic system is a combination of charged and neutral particles whose energy is not well predicted by theory.

Many liquid argon detectors use a time projection chamber (TPC) to measure the trail of ionization electrons left by a charged particle as it travels through the argon volume. Neutral particles, such as neutrons and neutrinos, escape detection in the TPC unless they interact with argon atoms and produce charged particles or excite a nucleus that can then emit a measurable signal. For charged particles, the trail of electrons is drifted to the wire planes of the TPC by an applied electric field. Impurities in the argon can capture the drifting electrons and skew the measurement of deposited charge left by the particle, causing a mis-reconstruction of the particle's energy. These effects must be taken into account to properly estimate the neutrino energy.

Recent studies\cite{Friedland:2018vry,PhysRevD.92.091301} have shown that neutrons can carry away a large fraction of the incoming neutrino energy, particularly for anti-neutrinos, and the secondary interactions can create a spray of small energy deposits in the TPC. Many of these energy deposits are below the detector threshold and thus are missed in the energy reconstruction. In order to estimate the true energy of the neutrino, the missing energy fraction must be well constrained. This requires a model for neutron interactions that is constrained by experimental data. In this letter we report a measurement of the neutron cross section on liquid argon using the Mini-CAPTAIN detector\cite{Berns:2013usa} for neutron kinetic energies between 100-800 MeV. 

\begin{figure}
\includegraphics[scale=0.4]{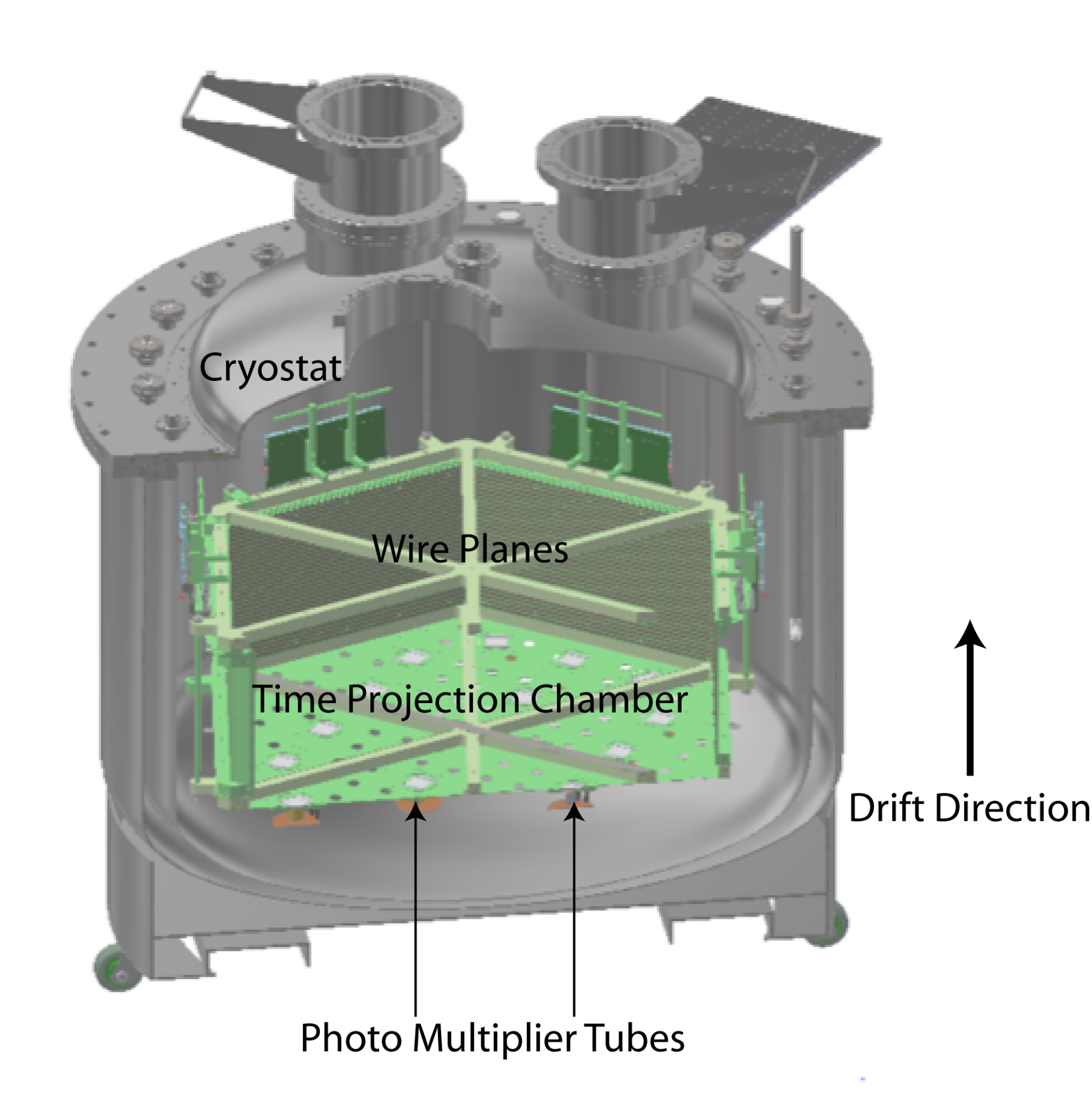}
\caption{Diagram of the Mini-CAPTAIN detector showing the cryostat, TPC, and selected Photo Multiplier Tubes. The top PMTs are not shown in the diagram.}
\label{fig:miniCAPTAIN_schematic}
\end{figure}

The Mini-CAPTAIN detector is a liquid argon TPC (LArTPC) with 400 kg of instrumented mass. In addition to the TPC, the detector is equipped with a photon detection system (PDS) consisting of 24 photomultiplier tubes (PMT) with 16 on the bottom of the TPC and 8 on the top (see Fig. \ref{fig:miniCAPTAIN_schematic}). The PDS measures the light from neutron interactions to establish an arrival time with respect to the beam spill time. The TPC is hexagonal in shape with an apothem of 50 cm and 32 cm of vertical drift between the cathode at the bottom and the anode at the top. The TPC has two induction planes (U and V) and a collection plane (X). The planes are made of copper beryllium wires 75 $\mu$m in diameter and spaced 3.125 mm apart. Each wire plane has 337 wires. The induction planes are rotated $\pm 60\degree$ with respect to the collection wire plane. The electric field applied corresponds to 500 V/cm, which results in an electron drift velocity of 1.6 mm/$\mu$s. The U and V planes are made transparent to electrons via biasing. The electron lifetime measured in the detector during the run was longer than 72.23 $\mu$s. This lifetime was long enough for tracks to drift to the collection plane given the measurements of our detector.

We deployed the Mini-CAPTAIN detector at the Los Alamos Neutron Science Center (LANSCE) in the 4FP15R beam line of the Weapon Neutron Research (WNR) facility. The facility provides a neutron beam with a broad energy spectrum up to an endpoint of 800 MeV.  Macropulses provide a 625 $\mu$s envelope in which individual sub-nanosecond bunches are spaced.  Our data-taking was completed with the beam and beam line in a special configuration.  The bunch spacing was 199 $\mu$s rather than the typical spacing of 1.8 $\mu$s.  The shutter to the beam line was partially closed to further reduce the neutron flux. These changes allowed us to provide a neutron flux of one neutron per macropulse to the detector - preventing event pile-up issues in the TPC where the drift time was 200 $\mu$s. The shutter configuration attenuated the low-energy portion of the neutron flux.

The data collected by the TPC were independent of the data from the PDS. The systems were later synchronized to produce a single data stream for analysis. Both systems received a signal from the beam facility for each spill, referred to as the radio-frequency pulse (RF) trigger. The TPC used this signal to trigger the data acquisition of 4.75 ms. This time corresponded to 1.85 ms of pre-trigger buffer data, 600 $\mu$s of beam time, and 2.3 ms of post-trigger data. These large windows of non-beam data were used to collect cosmic muon tracks in the detector for calibration. The PDS system also triggered on the RF signal and collected the light seen by the PMTs for 8 $\mu$s. The PDS could also trigger if enough light was seen by the PMTs independent of the RF signal. The PDS provided a more precise measure of the beginning of the drift time for the tracks in the argon than the time inferred from TPC data.

Raw signals from the TPC wires were filtered for electronics noise and were further processed to make hits from isolated peaks in the waveforms. Using a custom-made reconstruction algorithm, these hits were clustered to form 3D tracks. In each plane, hits located along straight lines are grouped together using a 2D Hough transform\cite{patent:3069654}. Additional hits were added to these track seeds if they were in a 40 mm box around the end of the track. This additional step improved the efficiency to reconstruct tracks that have gone through multiple scattering. A track candidate must have at least 3 hits from different wires to reduce uncorrelated hits making tracks. Track candidates were then associated with candidates in other planes. A 3D track was formed from a track candidate in the collection plane and at least one track candidate from an induction plane. We defined our coordinate system with the x axis along the beam direction and the z axis along the drift direction.

The primary purpose of the PDS is to measure the neutron arrival time at the detector. The neutron kinetic energy is determined by time-of-flight and is independent of the topology of the event as measured in the TPC. The timing information from the PDS also yielded a precise measurement of the vertical position of the related tracks in the TPC.
For all the PMTs, with 21 operational during our run, the analog waveforms were digitized with a 4 ns resolution. PMT hits were defined as peaks on the waveform amplitudes well above the noise pedestal.  Hits from the same PMT were combined if they were closely correlated inside a time window of 16 ns. 
Both photons and neutrons are produced at the beam target and produced scintillation signals visible to the PDS. The photons arrived in the detector at a characteristic time relative to the protons striking the target.  Using the known time structure of the beam, the relative time-of-flight of the photons and neutrons, and the distance between the target and the detector, the energy of each neutron interacting in the detector is calculated. The neutron kinetic energy resolution and the energy binning of the cross section is determined by the timing resolution of the PDS electronics.
Finally the PDS hits were associated to tracks in the TPC data if the times of the RF signals were within $\pm$100 ms of each other. Due to the better timing resolution in the PDS, multiple triggered PDS events were associated to the same TPC event.


We reconstructed 115,880 tracks from the 4.3-hour exposure at the WNR beam line. Candidate tracks must have at least one  associated PDS hit and the kinetic energy of the neutron that created the track must be at least 10 MeV. In addition, we required the tracks to be within a narrow time window of 31.25 $\mu$s from the RF signal. In order to further reduce the contamination from cosmic ray tracks in our sample, the starting position of the track must be at most $\pm$27 mm away from the beam line in the plane perpendicular to the drift direction. The fiducial volume for selected tracks was defined 3 wires away from the edges of the detector to be consistent with our reconstruction algorithm. Applying these cuts yields a selected sample of 9,911 tracks.

\begin{figure}
\includegraphics[scale=0.4]{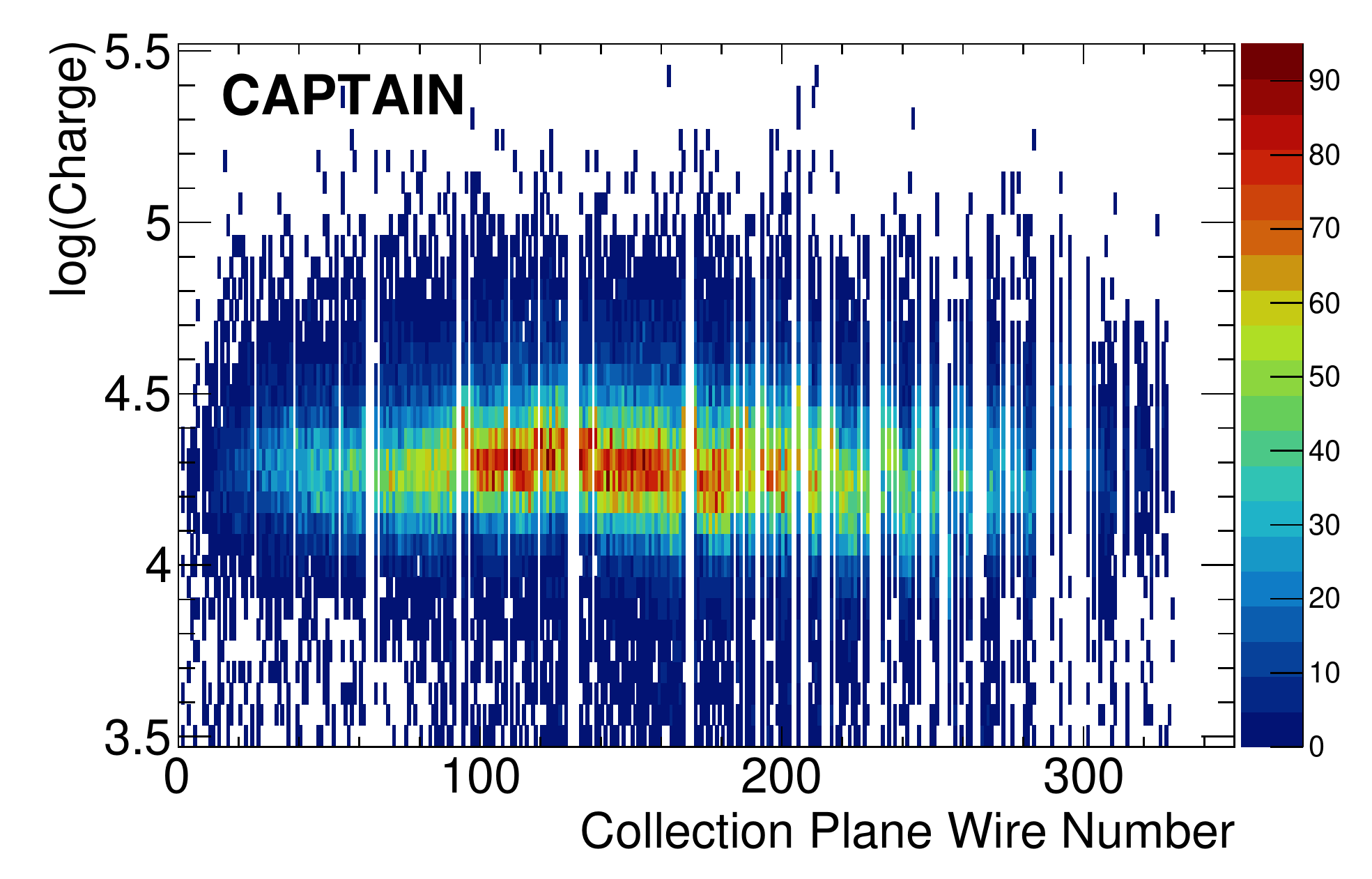}
\caption{Measured charge in collection plane for reconstructed cosmic ray tracks as a function of the wire number. The neutron beam comes from the right (large wire numbers). The gaps in the plot are wires that have been masked due to low efficiency.}
\label{fig:wire_inefficiency}
\end{figure}

Prior to analyzing candidate neutron tracks, the uniformity of the detector was studied using cosmic rays.  On the collection plane, which is used to directly determine the starting point of a track along the beam, 101 of 334 channels did not meet the data quality requirements. Of these channels, 18 (83) are in the downstream (upstream) region of the detector.  To mitigate the effect of the upstream inefficiency, tracks are reconstructed throughout the entire volume, but only those starting in the downstream region are considered in this analysis. Within the fiducial region, 89\% of the wires are active. Figure \ref{fig:wire_inefficiency} shows the number of electrons collected by each active wire in the collection plane for reconstructed cosmic ray tracks measured during the neutron exposure. The neutron beam entered the detector from the right side of the plot (higher wire numbers). The collected charge on each wire is uniform across the detector, and well above the charge threshold of 3000 collected electrons. A full simulation of the detector shows that the efficiency to reconstruct a track is uniform through-out the fiducial volume.  We reconstruct 2631 tracks as our final data set for the cross section measurement.

Our approach to measure the neutron cross section uses the fact that neutron flux decreases as a function of depth in the detector due to neutron interactions with argon.  The 
attenuation of the beam $dN_B/dx$ is proportional to the total neutron-argon 
cross section as seen in Equation \ref{eq:surv_prob}. The attenuation can be measured by choosing 
a particular event topology and measuring the change 
in the rate of this particular process as a function of depth in the 
detector.  
Provided that the fraction of the total cross section that 
results in the chosen event signature does not change as a 
function of depth, the measured change in its rate as a function 
of depth yields a measurement of the total cross section. Our chosen topology was single reconstructed tracks at least 15 mm long that started inside the fiducial volume.  
We make this measurement in narrow energy bins where the total 
cross section and partial cross sections are not expected to 
vary significantly.

The cross section can be extracted from the attenuation measurement using the formula:

\begin{equation}
dN_B/dx = -T\sigma_T N_B \quad \xrightarrow{} \quad N_B(x) = N_{0}e^{-T\sigma_{T} x},
\label{eq:surv_prob}
\end{equation}
where $T = \rho_{LAr}\times N_{\mathrm{Avogadro}}/m_{Ar}$ is the nuclear density of liquid argon, $\sigma_T$ is the total cross section and $x$ is the distance travelled in the material. There is a very small contribution from neutron-argon elastic scattering that could result in no visible signal and result in a 
neutron remaining in the beam.  Due to the expected small size of the elastic cross section for neutrons of 
kinetic energies above 100 MeV relative to other contributions 
to the total cross section\cite{neutronbookcrosssec}, it is not included in this analysis.

This approach to measure the cross section is valid under certain assumptions: The target volume must be uniform, the incoming flux must be monochromatic, the particles in the beam do not re-scatter within the selection region, and the additional neutrons from the background are negligible. These conditions apply in our analysis, either by construction or by our selection criteria. The argon target purity is an inherent requirement of the TPC to drift the electrons to the wire planes. Our beam was not monochromatic so we binned our tracks based on their kinetic energy determined by time-of-flight. To remove re-scatter events, we only consider events with a single track in them. Only 5$\%$ of events have multiple tracks in the selection region, which are removed from our analysis sample. Finally, by requiring that the tracks in the TPC are associated to a PDS hit triggered by the RF signal, we mitigate the external neutron background in our sample. Under these assumptions, we fitted the distribution of the track's starting position in the direction along the beam to an exponential function. We extracted the cross section in each bin from the fit and used the standard values for the nuclear properties of liquid argon ($T = (1.3973\ \rm{g/cm^3}\times6.022\times10^{23}\ \rm{n/mol}) / 39.948\ \rm{g/mol} = 2.11\times 10^{22}\ \rm{cm^{-3}}$).

\begin{figure}
\includegraphics[scale=0.4]{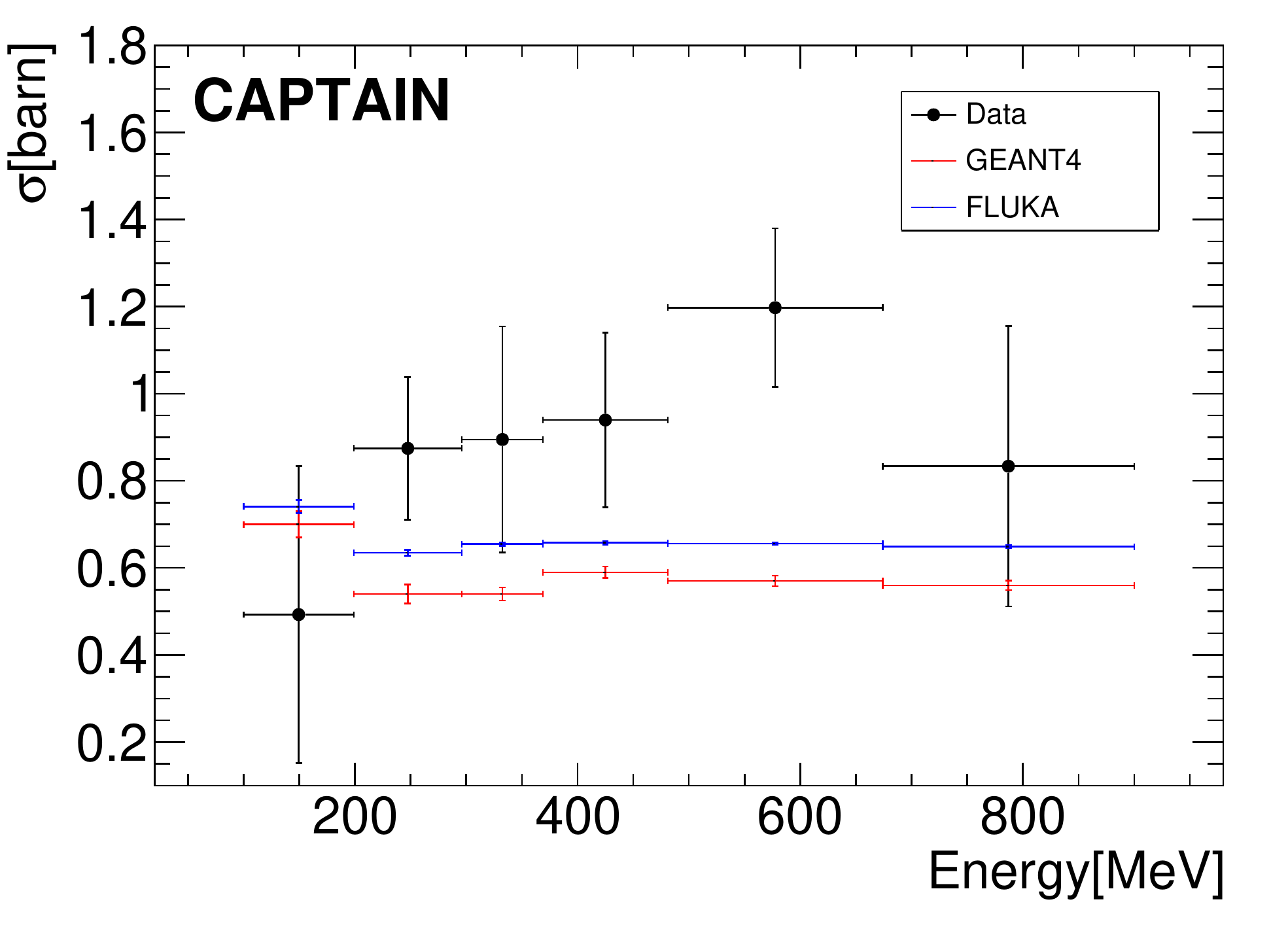}
\caption{Comparison of the measured neutron cross section as a function of neutron kinetic energy with values obtained from simulation. The cross sections are obtained from exponential fits to the distribution of tracks as a function of depth into the detector. The errors bars correspond only to the statistical uncertainty of the fits, for both data and simulation.}
\label{fig:neutronXsec}
\end{figure}

The cross section results are presented in Figure \ref{fig:neutronXsec} and more details about the fits are shown in Table \ref{tab:crosssec}. The bins chosen for the fits were determined by the procedure described in the literature\cite{binning}; therefore, the number of degrees of freedom is not the same for every fit. Most of the resulting fits have quite reasonable p-values consistent with the exponential fit assumption. The last energy bin has a rather large p-value of 0.998, but is still consistent with the exponential fit assumption given the limited statistics and the number of trials available.


A feature of this approach is that it does not depend on the normalization of the beam flux, but only on the relative intensities at different depths into the detector. Systematic effects associated with the overall detector performance are not relevant and only local fluctuations can impact the measurement. The uncertainty reported in the cross section measurements is dominated by the statistical uncertainty. 

\begin{table*}
\caption{Neutron cross section in bins of kinetic energy. The $\chi^2$ per degrees of freedom is presented, as well as the total number of tracks used for the fit in each bin. The exact functional form used for the fits is $f(x)=c_1e^{-c_2x}$. The cross section is extracted from the fitted parameter according to Equation \ref{eq:surv_prob}.}
\label{tab:crosssec}
\begin{ruledtabular}
\begin{tabular}{ c c c c} 
 Energy range [MeV] & Cross Section [barns] & $\chi^2$/ndof & Number of tracks\\
 \hline
100-199 & 0.49$\pm$0.34 & 1.48/3 & 264 \\
199-296 & 0.88$\pm$0.16 & 11.81/7 & 536 \\
296-369 & 0.89$\pm$0.26 & 4.739/5 & 329 \\
369-481 & 0.94$\pm$0.20 & 8.262/6 & 413 \\
481-674 & 1.20$\pm$0.18 & 5.713/6 & 624 \\
674-900 & 0.83$\pm$0.32 & 0.1323/4 & 252 \\
\end{tabular}
\end{ruledtabular}
\end{table*}

We developed a Monte Carlo simulation to study the performance of the reconstruction. The simulation is based on GEANT 4.10.3 with QGSP\_BERT physics list\cite{Agostinelli:2002hh} to model the interactions with the liquid argon and the simulation of the electronics is done with custom-made algorithms. We used our Monte Carlo simulation to constrain the size of the systematic uncertainties. In particular, we studied any effect that could distort the reconstructed position of our tracks. Using a simulated sample of protons with energies between 50 and 800 MeV, we compared the position of the reconstructed track with the true position. Any systematic bias in the position of the reconstructed tracks will distort the distributions in the fits. We determined our reconstruction algorithm resolution for the track position along the beam to be less than 20 millimeters in the downstream region of the detector. Given that the size of the bins in our fits is between 40-45 mm, these small position fluctuations can migrate at most 4$\%$ of our tracks. This is a negligible effect for the fits when extracting the cross section compared with the statistical uncertainty.

Another possible systematic effect in our reconstruction is the angle of the track. Given the orientation of the wire planes, a track that forms parallel to the wires will only make hits in just a few of them and might not be reconstructed by our algorithm. The simulated proton sample was generated in a forward cone along the beam direction. The reconstruction algorithm showed that there is not a particular angle for which we are less efficient. There is an ambiguity in determining the direction of a track in the TPC, as an upward forward moving particle looks just like a downward backward moving one. This effect is negligible in our data set since the tracks created by the striking neutrons are predominantly in the forward direction. 
Using the simulation, we were able to set an upper limit of a few percent on the systematic uncertainties and determined that they are negligible compared with the statistical uncertainty.

In addition to reconstruction systematic uncertainties, we also considered the effect of multiple-track events in the cross section calculation. 
To understand possible biases in the analysis, neutrons are simulated impinging on the detector at the measured coordinates of the neutron beam. After applying the analysis selection criteria to the simulated events, the cross section is extracted using the technique applied to data.
We compared the difference on the cross section calculation in events with exactly one track in them, as in our signal selection, and an inclusive sample of events with any number of tracks. The difference in cross sections between these two samples is on average less than 10$\%$. This effect is small and the total uncertainty in the cross section measurements are still dominated by the statistical uncertainty.

Figure \ref{fig:neutronXsec} shows a comparison of the measured neutron cross section with the values used by Monte Carlo generators, in this case GEANT4 and FLUKA\cite{fluka,fluka2}. 
The values from the generators presented here do not account for theoretical uncertainties in the models.
The measured cross sections are consistent with the previously reported data of other nuclear cross sections\cite{nitrogenOxygen,NSR1985CAZU}. 
It is also worth noting that the values extracted from the generators are consistent with the measured data and only small corrections are needed in the simulation.
The dominant source of uncertainty in our measurement is the limited statistics available for this analysis. The data set used for this analysis corresponded to the lowest intensity beam configuration of our run and in the future we will include additional data sets in other beam configurations.

In conclusion, we have presented the first measurement of the total neutron cross section on argon in the energy range of 100-800 MeV.  The measured cross sections are consistent with an energy averaged cross section of $0.91 \pm{} 0.10~\mathrm{(stat.)} \pm{} 0.09~\mathrm{(sys.)}~\mathrm{barns}$ with a $\chi^2 = 4.2/5\ \mathrm{ndof}$.
The measurements presented here will inform the developers of the simulation packages and constrain the uncertainties of the current models of neutron transport. In turn, this will improve the neutrino energy reconstruction performance of experiments attempting to resolve the CP violating phase and the neutrino mass hierarchy.

Research presented in this letter was supported by the Laboratory Directed Research and Development program of Los Alamos National Laboratory under project numbers 20120101DR and 20150577ER. This work benefited from the use of the Los Alamos Neutron Science Center, funded by the US Department of Energy under
Contract No. DE-AC52-06NA25396 and we would like
to thank Nik Fotiadis, Hye Young Lee and Steve Wender 
for assistance with the 4FP15R beamline.
We gratefully acknowledge the assistance of Mark Makela and 
the P-25 neutron team.
We further acknowledge the support of the US Department of 
Energy, Office of High Energy Physics and the University of 
Pennsylvania.
\bibliography{mybib}

\end{document}